\documentclass[floats,floatfix,showpacs,amssymb,prd,twocolumn,superscriptaddress,nofootinbib,nolongbibliography,reprint]{revtex4-1}
\usepackage{amssymb,amsmath,verbatim,mathtools,needspace,enumitem,etoolbox,graphicx,physics,microtype,afterpage,xspace,tabularx,lmodern,multirow}
\usepackage{gensymb}
\usepackage[dvipsnames, usenames]{xcolor}
\definecolor{linkcolor}{rgb}{0.0,0.3,0.5}
\usepackage[unicode, colorlinks=true, linkcolor=linkcolor, citecolor=linkcolor, filecolor=linkcolor, urlcolor=linkcolor, linktocpage, breaklinks]{hyperref}
\usepackage[all]{hypcap}
\usepackage[T1]{fontenc}
\usepackage[utf8]{inputenc}
\usepackage[usenames,dvipsnames]{xcolor}
\hypersetup{colorlinks=true,citecolor=romared,linkcolor=romared,urlcolor=romared}

\definecolor{romared}{RGB}{142,0,28}

\newcommand{\be}{\begin{equation}}
\newcommand{\ee}{\end{equation}}

\usepackage{enumerate}
\usepackage{tensor}
\usepackage{stmaryrd}
\usepackage[normalem]{ulem}
\usepackage{mathtools}
\usepackage{url}
\usepackage{multirow}
\usepackage{graphicx}
\usepackage{mathtools}
\usepackage{verbatim}
\usepackage{soul,xcolor}
\usepackage{subfig}
\setstcolor{red}

\def\be{\begin{equation}}
\def\ee{\end{equation}}
\newcommand{\beq}{\begin{eqnarray}}
\newcommand{\eeq}{\end{eqnarray}}

\begin{document}
\title{Towards VLBI Observations of Black Hole Structure}
\author{Raúl Carballo-Rubio}
\affiliation{CP3-Origins, University of Southern Denmark, Campusvej 55, DK-5230 Odense M, Denmark}
\author{Vitor Cardoso} 
\affiliation{Niels Bohr International Academy, Niels Bohr Institute, Blegdamsvej 17, 2100 Copenhagen, Denmark}
\affiliation{CENTRA, Departamento de F\'{\i}sica, Instituto Superior T\'ecnico -- IST, Universidade de Lisboa -- UL, Avenida Rovisco Pais 1, 1049-001 Lisboa, Portugal}
\author{Ziri Younsi} 
\affiliation{Mullard Space Science Laboratory, University College London, Holmbury St.~Mary, Dorking, Surrey, RH5 6NT, UK}

\begin{abstract}
Black holes hold a tremendous discovery potential.
In this paper the extent to which the Event Horizon Telescope and its next generation upgrade can resolve their structure is quantified.
Black holes are characterized by a perfectly absorptive boundary, with a specific area determined by intrinsic parameters of the black hole.
We use a general parametrization of spherically symmetric spacetimes describing deviations from this behavior, with parameters controlling the size of the central object and its interaction with light, in particular through a specular reflection coefficient $\Gamma$ and an intrinsic luminosity measured as a fraction $\eta$ of that of the accretion disc. This enables us to study exotic compact objects and compare them with black holes in a model-independent manner. We determine the image features associated with the existence of a surface in the presence of a geometrically thin and optically thick accretion disc, identifying requirements for VLBI observations to be able to cast meaningful constraints on these parameters, in particular regarding angular resolution and image dynamic range.
For face-on observations, constraints of order $\eta\lesssim 10^{-4}, \Gamma\lesssim 10^{-1}$ are possible with an enhanced EHT array, imposing strong constraints on the nature of the central object.
\end{abstract}

\maketitle

%%%%%%%%%%%%%%%%%%%%%%%%
\section{Introduction}
%%%%%%%%%%%%%%%%%%%%%%%%

The gravitational interaction is tested to exquisite precision in the weak-field regime~\cite{Will:2014kxa}. Gravitational-wave astronomy is now providing access also to strong-field regions, in violently dynamical regimes, and therefore to entirely new and exciting tests of Einstein's theory~\cite{Gair:2012nm,Berti:2015itd,Yunes:2016jcc,Barack:2018yly,Baibhav:2019rsa,Cardoso:2019rvt}. Thus far, gravitational-wave data is entirely consistent with all predictions of General Relativity (GR)~\cite{LIGOScientific:2020tif,LIGOScientific:2021sio}. 

It is, however, widely accepted that the description of gravity -- as provided by the classical equations of GR -- is incomplete. In particular, the theory breaks down in black hole (BH) interiors~\cite{Penrose:1964wq}. In addition, spacetime warping at BH horizons introduces puzzles in the semiclassical treatment of free fields~\cite{Unruh:2017uaw}. The resolution of these and other issues seems to require an improved theory, and corresponding different spacetime geometries. The scale and nature at which the new effects play a role is unknown, and could range from ``soft'' horizon scale changes in the geometry, to ``hard'' effects doing away with horizons completely, at least in a classical sense~\cite{Mathur:2009hf,Bena:2022ldq,Almheiri:2012rt,Berthiere:2017tms,Giddings:2022jda}.

It is up to observations to collect data pertaining to strong-field regions, in order to constrain the nature of dark compact objects~\cite{Cardoso:2019rvt}. Fortunately, BHs are also the simplest macroscopic objects in the cosmos, making the search for new physics associated with the absence of horizons particularly appealing~\cite{Cardoso:2019rvt,Cardoso:2017cqb}. 
Due to their very nature, it is impossible to ever have observational proof that horizons do exist~\cite{Abramowicz:2002vt,Cardoso:2019rvt}, but one can certainly {\it quantify} the statement that there is, or not, structure close to the Schwarzschild radius inducing deviations from a perfectly absorptive behavior. In the past, observations of supermassive BHs, in particular Sagittarius A*, were used to qualitatively push any putative surface to Planck distances away from the horizon~\cite{Broderick:2009ph,Broderick:2015tda}. The argument is based on thermodynamic equilibrium between the central object and its -- visible -- accretion disc. However, these arguments neglected strong lensing and conversion to other channels~\cite{Cardoso:2019rvt,Carballo-Rubio:2018jzw,Carballo-Rubio:2022imz}, and need to be revisited.

In addition to gravitational-wave astronomy, optical/infrared interferometry and mm-wavelength very large baseline interferometry (VLBI) are now able to scrutinize the region around BHs with unprecedented accuracy~\cite{EventHorizonTelescope:2019dse,GRAVITY:2020lpa,EHTC2022_PaperI}. Of special interest to us here is the groundbreaking Event Horizon Telescope (EHT) observations relating to images of BHs. Most notably, the Event Horizon Telescope $1.3$~mm wavelength images of the supermassive black holes in M87~\cite{EventHorizonTelescope:2019dse} and Sgr A*~\cite{EHTC2022_PaperI} have revealed bright, ring-like emission on scales of the Schwarzschild radius. Here, we wish to understand how these observations can be used to determine intrinsic features of BHs, such as the size of their boundaries, or their perfectly absorptive nature. This will give us information about the experimental designs that maximize the capabilities of testing these foundational aspects.

We use geometric units with the speed of light and Newton's constant $c=G=1$.

%%%%%%%%%%%%%%%%%%%%%%%%%%%%%%%%%%%%%
\section{The setup}
%%%%%%%%%%%%%%%%%%%%%%%%%%%%%%%%%%%%%
Obtaining images of alternatives to BHs is a necessary endeavour to test GR and the BH paradigm. A systematic study would ideally take into account several non-trivial aspects which, in practice, would involve including parameters and functional degrees of freedom~\cite{Carballo-Rubio:2018jzw} describing (i) the internal (bulk and surface) properties of the central object, (ii) the spacetime around them, and (iii) any matter propagating around and interacting with the central object.

Analyzing this complex question step by step, starting with the simplest possible models and gradually adding additional features, has the advantage of allowing us to examine the different physical aspects involved in a controlled manner. Moreover, simple models are interesting on their own right. Typically, their simplicity implies that the associated deviations from general relativity are less subtle than in more sophisticated models. Hence, simple models can be excellent markers of the boundary between observable and unobservable for a given VLBI design.

%%%%%%%%%%%%%%%%%%%%%%%%%%%%%%%%%%%%%%%%%%%%%%%%%%%%%%
\subsection{Parametrizing the central object}
%%%%%%%%%%%%%%%%%%%%%%%%%%%%%%%%%%%%%%%%%%%%%%%%%%%%%%

The simplest possible situation that we can analyze concerns spherically symmetric geometries with no additional external matter backreacting on the spacetime. In the context of ``hard'' but localized changes to the geometry, we can further make the assumption that the external geometry is given by the Schwarzschild solution,
\be
ds^2=-fdt^2+\frac{dr^2}{f}+r^2d\Omega^2\,.
\ee
In practice, these assumptions involve neglecting first the external and matter parameters (points ii and iii in the list above), focusing on the internal parameters of the central object. 

The assumption of spherical symmetry constrains the internal parameters of the central object: the size is uniquely determined by a single parameter, its radius $R$. We are mostly concerned with discerning structure at the horizon, and therefore looking for changes that occur on small scales. As such, we set the surface of the central object to be
\be
R=2M(1+\epsilon)\,,\label{eq:epsilon_def}
\ee
where we will be mostly interested in situations in which $\epsilon \ll 1$. Any coefficient describing bulk and surface properties must be a function of the radial coordinate only. 

Generating images involves the ray tracing of null rays. All the information that is needed to image a given central object boils down to characterizing the its interaction with light. This interaction can take place both at bulk and surface levels, although for light rays it is reasonable to assume that these interactions take place mostly at the surface level. We can ignore then for the moment the bulk parameters (which, in practice, is equivalent to assuming that the bulk is optically thick). In this simple situation, we just need three parameters: a radius for the central object $R$, a specular (or elastic) reflection coefficient $\Gamma$, and an intrinsic brightness ${\cal B}$ describing a locally isotropic surface emission due to a non-zero temperature (the latter is proportional to the dimensionless ratio between the re-emitted energy and incident energy, $\tilde{\Gamma}\propto\mathcal{B}$, used in~\cite{Carballo-Rubio:2018jzw}). This situation can be implemented by introducing a boundary in the Schwarzschild spacetime, thus creating a new spacetime with parameters $\{M,\epsilon,\Gamma,{\cal B}\}$, and modifying the propagation of null rays according to these coefficients.

In practice, we will thus be analyzing images of spacetimes with parameters $\{M,\epsilon,\Gamma,{\cal B}\}$, and comparing these with BH images. BH images are associated with the subset $\{M,\epsilon\ll 1,\Gamma=0,{\cal B}=0\}$.\footnote{A BH is strictly associated with $\epsilon=0$ but, as situations with $\epsilon\ll1$ contain deviations proportional to $\epsilon$ from a BH, the ubiquitous presence of numerical errors makes these indistinguishable in practice.}  As the nature of the central object can be very different when changing the values of the parameters involved, it is reasonable to expect that these images can be quite different. Both specular reflection on and intrinsic brightness at the boundary of the central object are nonexistent in BH spacetimes, which implies the possible existence of new image features intrinsically associated with these processes. We will see that this intuition is accurate, although whether these differences can be measured with a specific experimental setup is a more subtle question. While the Event Horizon Telescope collaboration has partially analyzed this issue for specific situations~\cite{EventHorizonTelescope:2022xqj}, our analysis is exhaustive regarding the parameter space described above.

%%%%%%%%%%%%%%%%%%%%%%%%%%%%%%%%%%%%%%%%%%%%%%%%%%%%%%%%%%%%%%%%%%%%%%%%%%%%%
\subsection{Interaction between the central object and the accreting material\label{subsec:interaction_disc_object}}
%%%%%%%%%%%%%%%%%%%%%%%%%%%%%%%%%%%%%%%%%%%%%%%%%%%%%%%%%%%%%%%%%%%%%%%%%%%%%
As the disc material falls onto the compact objects, its velocity as measured by locally static observers is increasing. In fact, one can calculate the center-of-mass energy with which it collides with material at the (static) surface. Take then a proton or electron in the accreting material of mass $m_0$ colliding with another one at the surface. The center of mass energy is
\beq
E_{\rm CM}&=&m_0\sqrt{2}\sqrt{1-g_{\mu\nu}u^\mu_{(1)}u^\nu_{(2)}}\\
&\approx& \sqrt{2E}\frac{m_0}{\epsilon^{1/4}}\,.
\eeq
where $u^\mu_{(1)},u^\mu_{(2)}$ are the 4-velocities of the particles~\cite{Problem_book}. In the last step we assume that one particle sits at the surface of the central object, while the other is described by a conserved energy parameter $E$. For a particle that falls from the inner edge of the accretion disc, presumably at the innermost stable circular orbit, $E=2\sqrt{2}/3$.

We see that for $\epsilon \lesssim 10^{-10}$, the center of mass energy for proton-proton collisions goes beyond the electroweak scale $\sim 200$ GeV. The collision by-product therefore consists of all possible Standard Model particles compatible with the scattering process. Since for small $\epsilon$ all the by-products are strongly lensed and made to interact again with the central object, it is reasonable to expect that it thermalizes, reaching equilibrium with all the Standard Model species: our central object behaves essentially as a Hawking-radiating BH~\cite{Page:1976df,Giddings:2001bu,Kanti:2004nr,Cavaglia:2002si}, although at a temperature dictated by the feeding disc. For small enough $\epsilon$, the CM collisions occur at high energy. For objects with effective temperatures down to of order 100 GeV, all Standard Model particles may be treated as essentially massless whereas for temperatures smaller than this there will be phase space suppression for the heavy gauge bosons and top quarks. In conclusion, we argue that the ultracompact object should be re-emitting a nearly isotropic radiation, which is composed of $\sim 1\%$ of photons, the rest are neutrinos, electrons etc. 

The previous argument strongly suggests that ultracompact central objects will be dimmer than previously thought in the electromagnetic band (and brighter in the neutrino band for example). 
A more rigorous analysis needs to be done to decide on the fate of heavier species (which may be unable to exit the central object due to the large gravitational potential). In fact, for cold (as measured at infinity) central objects with $kT_\infty \lesssim mc^2$, an elementary particle of mass $m$ is unable to reach large distances and must fall back. In fact, for Sgr A*, EHT infrared observations and the assumption that the object emits blackbody radiation indicate that the central object is colder than $\lesssim 10^4$ K~\cite{EventHorizonTelescope:2022xqj}. In this case, only photons and neutrinos reach infinity and therefore the fraction of accreting energy coming out as photons increases from $1\%$ to a sizeable fraction of the incoming flux~\cite{Broderick:2007ek}.\footnote{We thank Ramesh Narayan for bringing the Sgr A* constraints to our attention~\cite{EventHorizonTelescope:2022xqj}.}

Recent work on the appearance of horizonless geometries assumed that the surface of the central object emits blackbody radiation at the same temperature as the inner accretion flow $8\times 10^{10}$ K~\cite{Vincent:2020dij}. This assumption is, in our opinion, ad hoc and lacks a sound theoretical basis. Indeed, were the central object + accretion disc system in thermal equilibrium, the locally measured (Tolman) temperature should be~\cite{Tolman:1930zza,Santiago:2018kds}
\be
T(r)=\frac{T_0}{\sqrt{f}}\,,
\ee
expressing the fact that heat is also subjected to gravity. This law would predict a hotter central object. The ensuing radiation, after redshifts caused by the gravitational potential, would give energy conservation: what comes in from the disc, comes out as radiation. In other words, as is clear from Eq. (7) in Ref.~\cite{Vincent:2020dij}, the more compact the central object, the colder it is. Thus it is far from equilibrium with the accretion disc. 

%%%%%%%%%%%%%%%%%%%%%%%%%%%%%%%%%%%%%%%%%%%%%%%%%
\subsection{Ray-tracing and accretion flow model}
%%%%%%%%%%%%%%%%%%%%%%%%%%%%%%%%%%%%%%%%%%%%%%%%%
Calculating images of the accretion disc and reflecting surface requires determining the trajectories of photons (null geodesics) within the background spacetime geometry.
In this work we use the general-relativistic radiative transport code \texttt{BHOSS}~\cite{Younsi2012,Younsi2016,Younsi2021}.
The geodesic equations of motion are formulated as a system of eight coupled first-order ordinary differential equations
\begin{equation}
\dot{x}^{\mu} = k^{\mu} \,, \qquad \dot{k}^{\mu} = - \Gamma^{\mu}_{\alpha\beta} k^{\alpha} k^{\beta} \,,
\label{eqn_geo}
\end{equation}
where $x^{\mu}$ denotes the position 4-vector, $k^{\mu}$ the (contravariant) photon 4-momentum, $\Gamma^{\mu}_{\alpha\beta}$ the Christoffel symbols, and an overdot denotes differentiation with respect to the affine parameter.
Equations \eqref{eqn_geo} are integrated numerically in \texttt{BHOSS} using an adaptive fourth order Runge-Kutta-Fehlberg method (with fifth order error estimate).
The integration tolerance is generally set to $\sim 10^{-10}$, sufficient for the purposes of this work.
Details regarding the geodesic initialisation may be found in \cite{Younsi2016}.

We consider the source of illumination to be a geometrically-thin and optically thick equatorial accretion disc.
The disc is formally infinite in extent, with an inner edge specified by the radius of the innermost stable circular orbit (ISCO).
The accretion disc material motion is assumed Keplerian, with four velocity, $u^{\mu}$, given by~\cite{Bardeen1972}
\begin{equation}
u^{\mu} = u^{t} \left( 1,\, 0,\, 0,\, \Omega_{\rm K} \right) \,,
\end{equation}
where $u^{t}$ and the Keplerian angular velocity, $\Omega_{\rm K}$, are
\begin{equation}
u^{t} = \frac{r^{1/2}}{\sqrt{r - 3M}} \,, \qquad \Omega_{\rm K} = \frac{\sqrt{M}}{r^{3/2}} \,.
\end{equation}
Once $\left(x^{\mu},\,k^{\mu}\right)$ are calculated for a  photon, the energy correction factor is calculated as $g=-k_{\mu}u^{\mu}$.
The stationary and axisymmetric nature of the spacetime geometry gives rise to two Killing vectors, from which one obtains $k_{t}\equiv -E$ and $k_{\phi}\equiv L_{\rm z}$, where $E$ and $L_{\rm z}$ define the energy and angular momentum (along the spin axis) of the photon, respectively.
The energy correction factor may then be written as $g=u^{t}\left(E - \Omega_{\rm K} L_{\rm z}\right)$.

Finally, one must prescribe a local emissivity for the accretion disc.
Since the disc is planar and opaque, we can neglect the effects of absorption and accumulation of intensity along the ray, thereby simplifying the problem of radiative transfer to determining the intersection of each ray with the disc surface, wherein each ray is prescribed a local monochromatic intensity, $j$, which is weighted by $g$.
In this work we adopt the profile $j(r)\propto r^{-n}$ with $n=3$, where $r$ is the disc's radial co-ordinate and $n$ has been chosen to ensure a finite total flux from the disc.

%%%%%%%%%%%%%%%%%%%%%%%%%%%%%%%%%%%%%%%%%%%%%%%%%%%%%%%%%%%%%%%
\section{Image features associated with specular reflection
\label{sec:cold_eco}
}
%%%%%%%%%%%%%%%%%%%%%%%%%%%%%%%%%%%%%%%%%%%%%%%%%%%%%%%%%%%%%%%
%
\begin{figure}
%\centering
\includegraphics[width=0.4\textwidth]{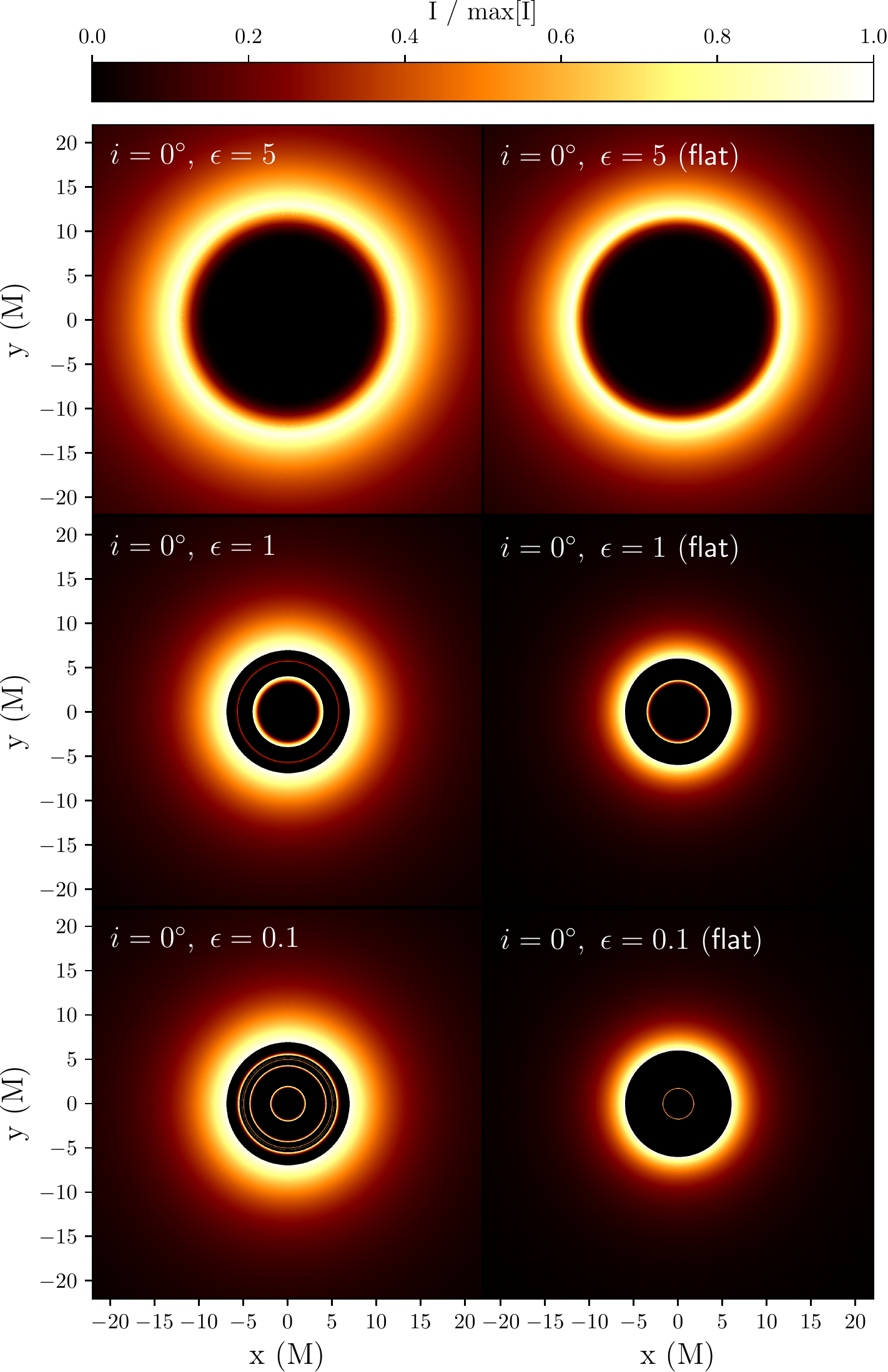}
\caption{Images of spacetimes with and accretion disc and a spherical mirror in Schwarzschild (left) and flat (right) geometries, with $i=0^\circ$ and $\epsilon=5, 1,\, 0.1$, in that order, from top to bottom. Note that the brightness of the accretion disc is noticeably different for the Schwarzschild and flat spacetimes, as the emission profile of the disc adjusts to the background gravitational field. Also, images for flat spacetimes do not have an associated mass scale, but we are still using the same field of view to simplify the comparison between both situations. Note also the distinctive features of curved spacetimes, which give rise to multiple rings, whereas a mirror in flat space only provides a single ring, from specular reflection.
\label{fig:image_inclination1}
}
\end{figure}
\begin{figure}
%\centering
\includegraphics[width=0.4\textwidth]{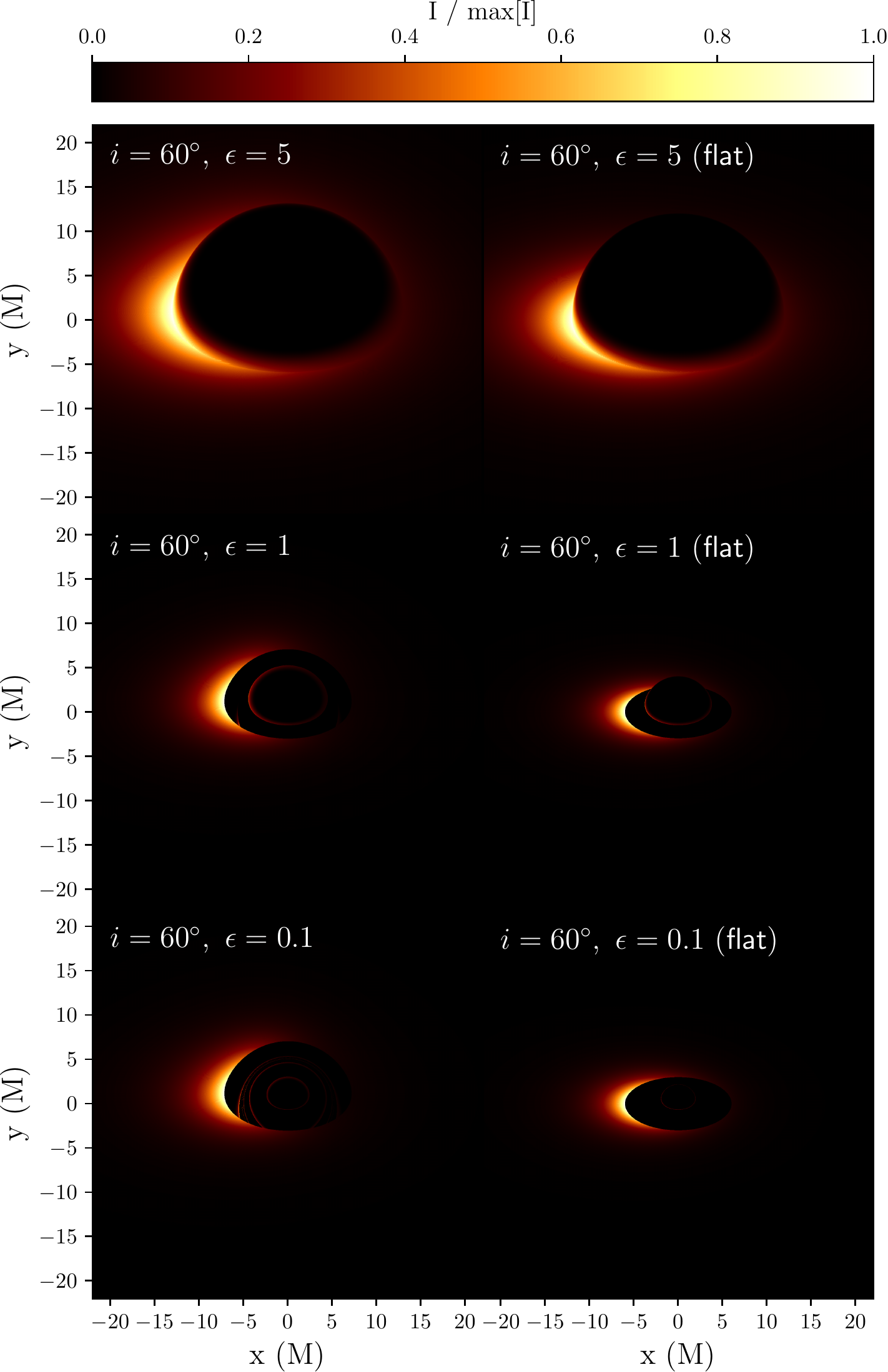}
\caption{Images for spherical mirrors in Schwarzschild (left) and flat (right) spacetimes, with $i=60^\circ$ and  $\epsilon=5, 1,\, 0.1$, 
in that order, from top to bottom.}
\label{fig:image_inclination60}
\end{figure}
\begin{figure*}
\centering
\includegraphics[width=0.99\textwidth]{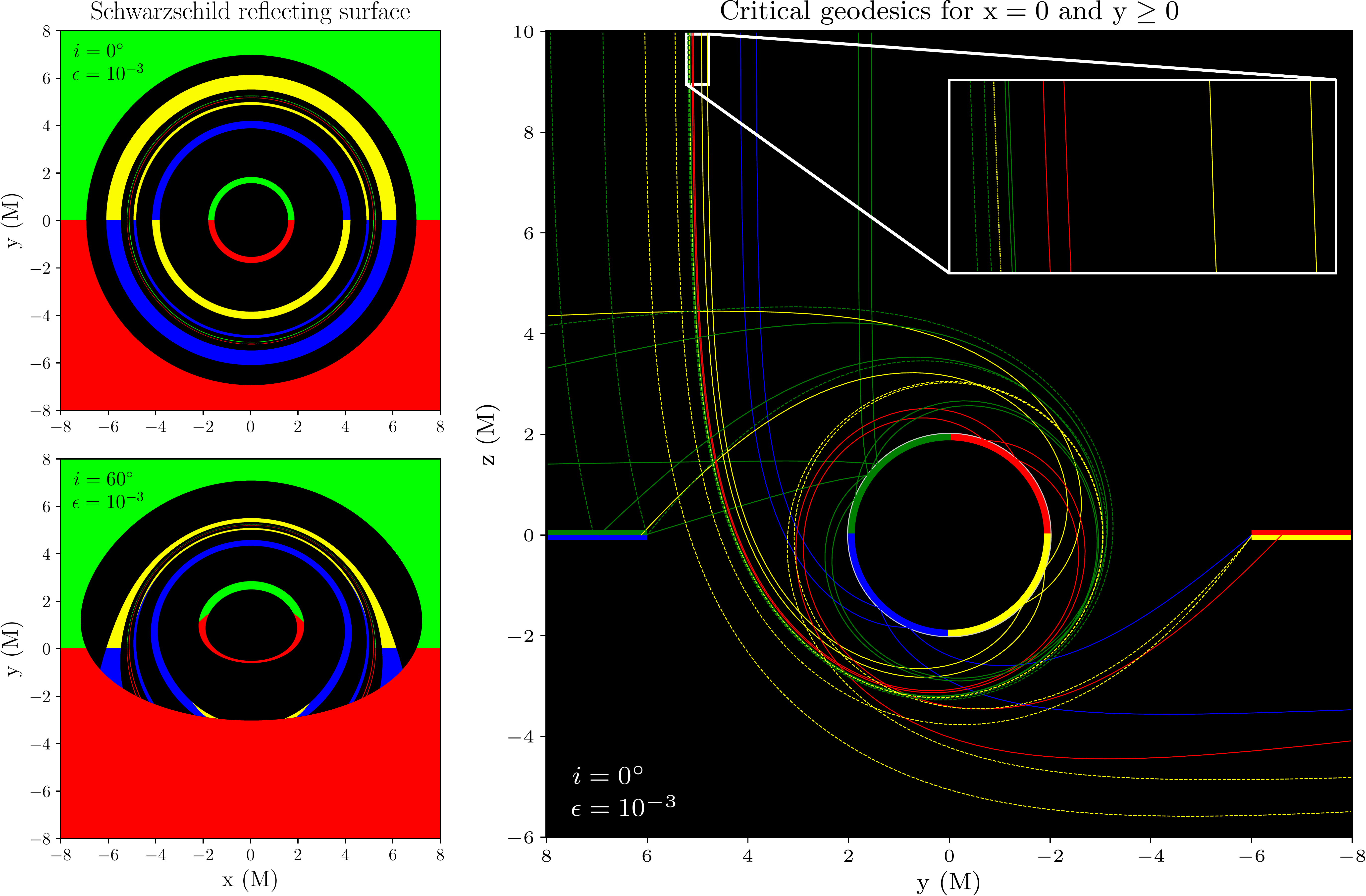}
\caption{
Origin of image geodesics for Schwarzschild reflecting surfaces for a face-on observer, i.e., $i=0^{\circ}$ (upper left) and $i=60^{\circ}$ (lower left). Pairs of critical geodesics which constitute the inner and outer boundaries of the ``photon rings'' and ``reflected rings'' in the upper left panel, as seen therein along the axis $\mathrm{x}=0$ and $\mathrm{y}\ge 0$. The observer is located at $\mathrm{z}\simeq 10^{4}~M$.
Reflected photons are denoted by solid lines and coloured by the quadrant of the surface they encounter. These photons comprise the reflected rings. Photons which are not reflected are denoted by dashed lines and coloured by the region of the accretion disc from which they are produced. The zoom-in view shows that the $n=3$ photon ring (dashed yellow) and the $m=4$ reflected ring (solid green) are in close proximity to one another but do not overlap.}
\label{fig:i0_geodesic_origins}
\end{figure*}
Given that the generation of images is a linear process, we can analyze the consequences of having a reflecting surface, or one with an intrinsic brightness independently.

Hence, for the moment the only relevant physical process will be reflection, and we will focus on the case $\Gamma=1$ that maximizes the associated features. Lower values of $\Gamma$ would just make these features dimmer, as we will discuss later. We have generated a series of images for different values of the radius $R$, parametrized in terms of a dimensionless parameter $\epsilon$ as defined in Eq.~\eqref{eq:epsilon_def}. Our results for the images of horizonless objects when illuminated by the accretion flow described above are summarized in Figs.~\ref{fig:image_inclination1}-\ref{fig:image_inclination60}. The observer is located at an angle $i$ with respect to the disc. For $i=0^{\circ}$ the observer is at the poles, seeing the disc head-on. For $i=90^{\circ}$ the observer is on the disc plane.

These images display rich features associated with different trajectories of light. However, not all these features are present for all values of $\epsilon$. In fact, images become increasingly complex as $\epsilon$ decreases. It is thus useful to start discussing the features in the simpler images obtained for large values of $\epsilon$, and then discuss the new features that appear successively as the value of this dimensionless parameter decreases. The details of these images can be understood with the help of Fig.~\ref{fig:i0_geodesic_origins}, which traces the origin of geodesics reaching the observer, and can be grouped as follows.

%%%%%%%%%%%%%%%%%%%%%%%%%%%%%
\subsection{Specular ring}
%%%%%%%%%%%%%%%%%%%%%%%%%%%%%
Specular reflection at $r=R$ results in a ``specular'' ring, associated with the specular (or elastic) reflection in the surface of the photons emitted from the accretion disc. This feature is in fact present for all values of $\epsilon$, as long as there is an accretion disc around the central object.

The largest value that we have considered for the images is $\epsilon=5$, for which $R=12M$ is twice the radius of the innermost stable circular orbit (ISCO). In this case, the obtained images are a superposition of the direct and reflected images of the accretion disc. Moreover, due to the large value of $\epsilon$, differences between images obtained for the Schwarzschild and Minkowski spacetimes are negligible in this case (see Figs.~\ref{fig:image_inclination1}-\ref{fig:image_inclination60}).

Figure~\ref{fig:image_inclination1} also shows that, as $\epsilon$ decreases, the direct and reflected images of the accretion disc become separated, while also being possible to clearly see the effect that gravitational lensing has on the specular ring for the Schwarzschild spacetime. Note that the position of the specular ring changes noticeably with $\epsilon$, as expected from its interpretation.

We are including plots of the Minkowski case in this section as a check of the sensibility of our results. The problem of reflection from a sphere is well studied in flat space~\cite{Berry:1923,2017arXiv170306768K}. For an observer looking face on from large distances, points in the (thin) accretion disc at a distance $r$ from the surface of the sphere will be seen to come from a ring of radius $L$ with
\beq
L&=&R\cos{\left(\pi/4+\arctan\Upsilon\right)}\,,\\
\Upsilon&=&\frac{1}{2}\left(1-\sqrt{1+8x}+\sqrt{2}\sqrt{-1+4x-\sqrt{1+8x}}\right)\,,\\
x&=&\frac{r^2}{R^2}\,.
\eeq
For $R=2M$ and $r=6M$ for example, one finds $L\simeq1.591M$. The end of the accretion disc (i.e., points at infinity) corresponds to points $x\sim 0$, and a ring at $L=\sqrt{2}/2M\sim 1.414M$, thus overall one gets a bright circular ring of thickness $0.176M$. The results above can be extended to an arbitrary observation inclination angle~\cite{Berry:1923,2017arXiv170306768K}. It is interesting that the projected ring has maximum thickness not for a head-on observer ($\theta=0$), but for an inclination $\theta\sim 8.5$ degrees.

%%%%%%%%%%%%%%%%%%%%%%%%%%%%%%%%%%%%
\subsection{Lensing/photon ring}\label{sec:lensing_photon_ring}
%%%%%%%%%%%%%%%%%%%%%%%%%%%%%%%%%%%%

The only role that gravitational lensing plays in the images discussed so far is the distortion (relative to the Minkowski case) of the specular reflection of the accretion disc. However, as $\epsilon$ decreases, gravitational lensing becomes large enough that photons from the accretion disc can circle the spherical mirror from below and then hit the observer without ever touching the surface of the central object, as shown clearly also in Fig.~\ref{fig:i0_geodesic_origins}. This leads to a secondary image of the accretion disc. One can calculate the minimum radius beyond which this secondary ring appears. The behavior of null geodesics was studied in detail by Chandrasekhar~\cite{MTB} (see also~\cite{Iyer2007}). The lensing deflection angle of a photon can be expressed in terms of elliptic integrals. In particular, if we define the perihelion distance $P$, then the lensing angle $\Theta$ of a photon in the Schwarzschild background is:~\footnote{To keep in line with standard definitions -- and in contrast with Chandrasekhar, we adopt the notation where the argument of the elliptic functions is $k^2$ instead of $k$.}
\begin{subequations}
\begin{eqnarray}
Q &\equiv& \sqrt{(P - 2M)(P + 6M)} \,, \\
k^{2} &\equiv& \frac{Q - P + 6M}{2Q} \,, \\
\chi_{\infty} &:=& \arcsin{\sqrt{\frac{Q - P + 2M}{Q - P + 6M}}} \,, \\
\Theta &:=& 4\sqrt{\frac{P}{Q}}\left[K(k^{2})-F(\chi_{\infty},k^{2})\right] - \pi \,.
\end{eqnarray}
\end{subequations}
One can easily calculate the critical radius for which the bending angle is $\pi/2$ or larger, corresponding to photons from distant regions of the accretion disc being seen by the face-on observer. The critical radius is $R\simeq 4.65958M$ ($\epsilon\simeq 1.32979$).

\begin{figure}
\includegraphics[width=0.3\textwidth]{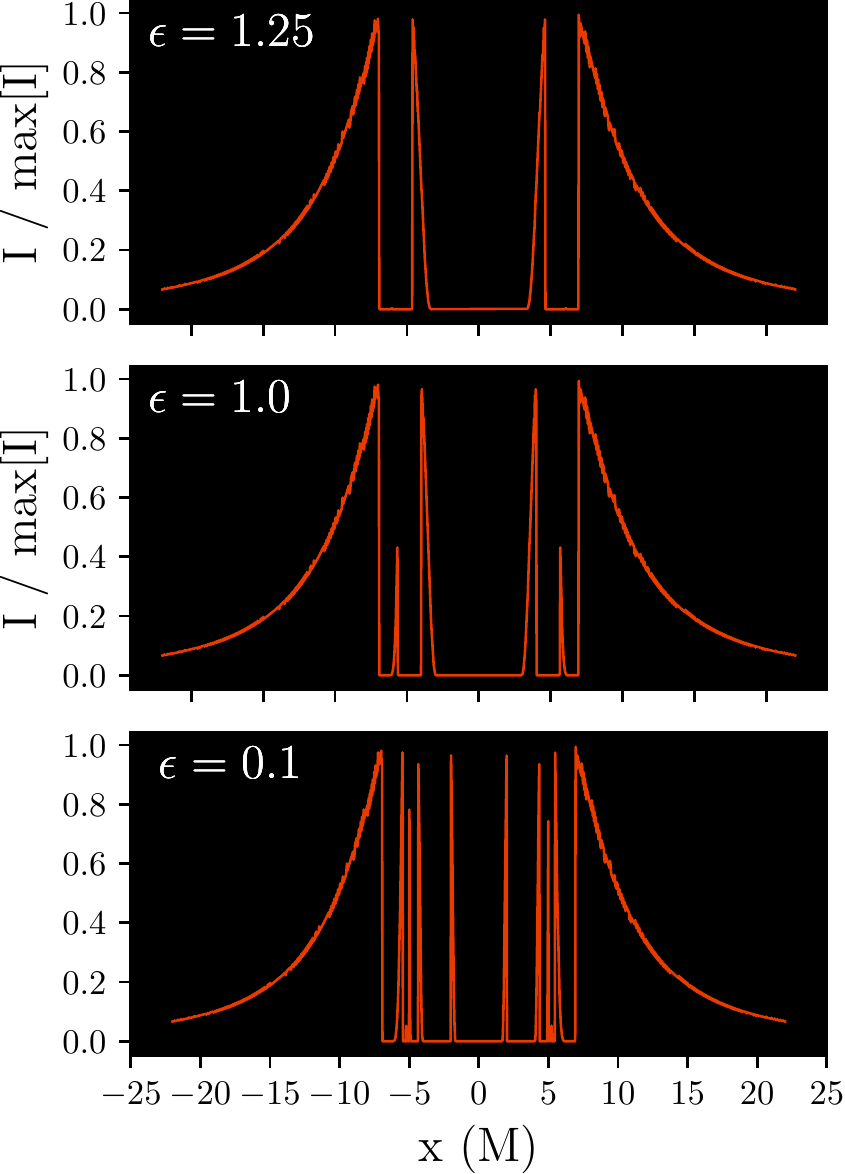}
\caption{Cross-section of images for $\epsilon=1.25$, $\epsilon=1.0$ and $\epsilon=10^{-1}$, respectively ($i=0^\circ$). The lensing/photon ring starts to be visible for $\epsilon=1.25$ (note the small peaks in between the accretion disc and the specular ring in the intensity profile), and it becomes brighter for decreasing values of $\epsilon$. As $\epsilon$ decreases, there also appear additional rings between the lensing/photon and the specular rings.\label{fig:sequence}}
\end{figure}

This analysis of null geodesics suggests that, for $\epsilon\lesssim 1.33$, there should be an additional ring in the corresponding images, which we call ``lensing/photon'' ring (additional details on this name below). Note that the critical value $\epsilon\simeq1.33$ indicates the threshold below which a distant observer is able to receive at least one photon that has circled the spherical mirror. However, for the lensing/photon ring to be manifest in images, enough photons from the accretion disc must circle the spherical mirror. Hence, it is reasonable to expect that, depending on the resolution used to generate and analyze the images, as well as the emission profile of the accretion disc, the lensing/photon ring should start to be visible for a value of $\epsilon$ slightly below this threshold. This is compatible with our images (Fig.~\ref{fig:sequence}). 

As shown in Fig.~\ref{fig:sequence}, we start seeing the lensing/photon ring for $\epsilon= 1.25$. For $\epsilon=1.25$ the lensing/photon ring is quite dim. This changes if we consider smaller values of $\epsilon$, as the lensing/photon ring becomes brighter as the compactness increases, until this brightness hits a plateau for $\epsilon\leq0.5$. However, the position of the lensing/photon ring does not change with $\epsilon$. This suggests that this ring is associated with the geometry around and outside the spherical mirror, but not with reflection. To see this explicitly, we generated images with perfectly absorptive boundary conditions, verifying that such lensing ring is still present (see also Fig.~\ref{fig:i0_geodesic_origins}).

Regarding the name lensing/photon ring, we are partially following the notation in Ref.~\cite{Gralla:2019xty}. It is important to keep in mind that a general image (of a BH or any other object) does not have a single ring, but a sequence of them. What is generally called ``photon'' ring actually refers to a ring structure that shows different features associated with different photon orbits that can circle the central object multiple times. The authors of~\cite{Gralla:2019xty} proposed a specific way to split these features, further distinguishing between the lensing and photon rings, the first one being associated with photons that circle the central object once, thus covering a total angle of $2\pi$, while the second one is associated with photons that at least cover an angle of $5\pi/2$ and thus can circle the central object an unlimited amount of times. 

For our purposes, the most crucial aspect to keep in mind is that not all the features of the lensing/photon ring appear for the same values of compactness. In fact, in the terminology of~\cite{Gralla:2019xty}, our results show that the features associated with the lensing ring appear for values of the radius greater than $3M$ ($\epsilon=0.5$), in particular for $R\lesssim4.6$ ($\epsilon\lesssim1.3$). We have discussed previously that, for this critical value of the radius, photons can start to circle around the spherical mirror and still reach a distant observer. As $\epsilon$ decreases, photons can complete additional orbits around the spherical mirror before reaching a distant observer, which means that additional photon orbits contribute to the lensing/photon ring. As a result, the lensing/photon ring gradually becomes as bright as the one in a BH image, and in fact becomes indistinguishable from the latter for $\epsilon=0.5$. 

In conclusion,

\begin{itemize}
\item The existence of a photon ring in an observational image is a relatively loose indicator of the compactness of the object, which can be anywhere in the interval $\epsilon\lesssim1.3$ ($R\lesssim4.6$).

\item More refined evaluations of the intensity profile and substructure of the photon ring are needed to argue to further constrain the radius of the central object down to $\epsilon\lesssim0.5$ ($R\lesssim3.0$).
\end{itemize}

%%%%%%%%%%%%%%%%%%%%%%%%%%%%%%%
\subsection{Reflected rings}
%%%%%%%%%%%%%%%%%%%%%%%%%%%%%%%
Aside from the specular and lensing/photon rings discussed above, the spherical mirror leads to the appearance of additional rings associated with reflection, as can be seen for instance in Fig.~\ref{fig:sequence}. Recall that the specular ring is also associated with reflection, as it is the direct specular image of the accretion disc. On the other hand, the lensing/photon ring is associated with photon orbits that circle the central object, without touching the latter (thus never being elastically reflected). The reflected rings are associated with photon orbits that both circle the central object but have also been elastically reflected in the latter.\footnote{Our numerical studies indicate that for $\epsilon>0.5$ (i.e., outside the unstable photon orbit radius), the photon rings and reflected rings occur in pairs. Following Sec.~\ref{sec:lensing_photon_ring}, there are no photon rings ($n=0$) and only the specularly reflected ring (hereafter $m=0$) when $\epsilon \gtrsim 1.3298$. For $\epsilon \gtrsim 0.7603$, the $n=1$ photon ring and an $m=1$ reflected ring are present, and this pattern continues as $\epsilon\rightarrow 0.5$. The effect of decreasing epsilon is to reduce the diameter and thickness of all reflected rings, whilst leaving the geometry of the photon rings unaffected. See also Fig.~\ref{fig:i0_geodesic_origins}.}

\begin{figure}
\subfloat{\includegraphics[width=0.4\textwidth]{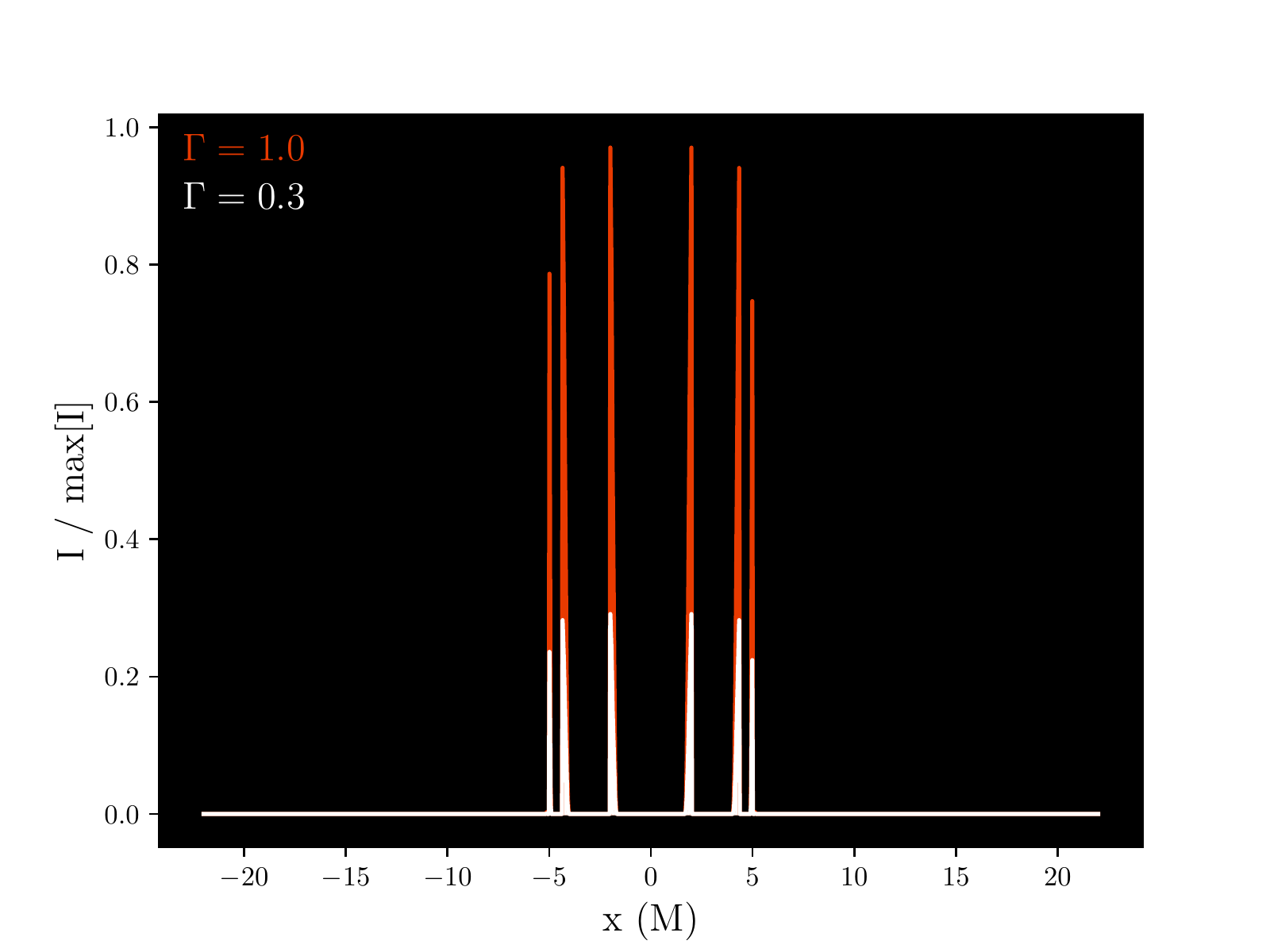}}
\caption{Profile that results from subtracting the cross-section for absorptive boundary conditions to the cross-section for reflective boundary conditions, for $i=0^\circ$, $\epsilon=10^{-1}$ and different values of $\Gamma=1$ (orange) and $\Gamma=0.3$ (white). The features shown are associated with specular reflection and their size scales linearly with $\Gamma$, with the innermost ring corresponding to the specular reflection of the accretion disc. \label{fig:int_ref-abs}}
\end{figure}

To isolate the features associated with reflection, we can subtract the intensity profile for reflective and absorptive boundary conditions. Fig.~\ref{fig:int_ref-abs} shows the result of this procedure for $\epsilon=10^{-1}$.

%%%%%%%%%%%%%%%%%%%%%%%%%%%%%%%%%%%%%%%%%%%%%%%%%%%%%%%%%%%%%%%
\section{Image features associated with re-emission or intrinsic brightness
\label{sec:hot_eco}
}
%%%%%%%%%%%%%%%%%%%%%%%%%%%%%%%%%%%%%%%%%%%%%%%%%%%%%%%%%%%%%%%
%%%%%%%%%%%%%%%%%%%%%%%%%%%%%%%%%%%%%%%%%%%%%%%%%%%%%%%%%%%%%%%

So far we have been analyzing situations in which the central object has a vanishing intrinsic brightness ${\cal B}=0$ due to it being extremely cold. As we discussed above in Sec.~\ref{subsec:interaction_disc_object}, the interaction between radiation and infalling material from the disc and the central object will heat the central object and lead to emission of all standard model particles. In addition, specular reflection (the channel we have studied so far) is only good as long as the wavelength of the radiation is larger than the typical particle separation in the scattering surface. In other words, the surface of the reflecting object must be smooth on a wavelength-scale. Hence, including re-emission under the form of an intrinsic brightness is physically well-motivated.

Fortunately, the linear nature of the problem simplifies the analysis. From the perspective of images, a nonzero value of ${\cal B}$ translates into the existence of a filled circle, with its brightness being a function of this parameter. It is possible to treat ${\cal B}$ as a completely independent parameter, to be constrained by observations. However, it is useful to relate ${\cal B}$ to the total flux radiated from the accretion disc, as described in the section below.

%%%%%%%%%%%%%%%%%%%%%%%%%%%%%%%%%%%%%%%%%%%
\subsection{Flux from the central object\label{flux_luminous}}
%%%%%%%%%%%%%%%%%%%%%%%%%%%%%%%%%%%%%%%%%%%

The accretion disc has a local emissivity profile $j \propto r^{-n}$. We can formally calculate the total flux radiated from the accretion disc as:
\begin{eqnarray}
F_{\rm tot} &\propto& 2 \iint\limits_{\rm Disc} j(r) \, \mathrm{d}r \, \mathrm{d}\phi = \int_{r_{\rm ISCO}}^{\infty} j(r) \, 4\pi r \, \mathrm{d}r \nonumber \\
&=& \frac{4\pi \, (r_{\rm ISCO})^{2-n}}{n-2} \,,
\end{eqnarray}
where the factor of $2$ accounts for emission from both the upper and lower parts of the disc.
In this work we assume the background spacetime is Schwarzschild, yielding $r_{\rm ISCO}=6M$, and the index $n=3$.
The dimensionless total flux of the accretion disc in this work is therefore $F_{\rm tot}\propto 2\pi/3$.
Since we omit absorption and are only concerned with emission at the disc surface, the constant of proportionality is arbitrary and hereafter fixed to unity.

In considering the emission from the reflecting surface, we apportion a fraction, $\eta$, of the total disc flux to the surface of the sphere, $F_{\rm sphere} = \eta \, F_{\rm tot} = 2\pi \eta/3$. From this, we uniformly distribute the aforementioned flux over the reflecting surface, yielding the local emissivity of the reflecting surface as:
\begin{equation}
j_{\rm surf} = \frac{\eta}{6 R^{2}} \,.
\label{surface_emissivity}
\end{equation}
Note that Eq.~\eqref{surface_emissivity} specifies that the local emissivity of the reflecting surface decreases (i.e., it becomes dimmer) as its radius increases, as expected.
The reflecting surface is assumed to be stationary and hence the local energy correction factor is unity and the flux may simply be added to that from the accretion disc. 

\begin{figure}
\centering
\includegraphics[width=0.49\textwidth]{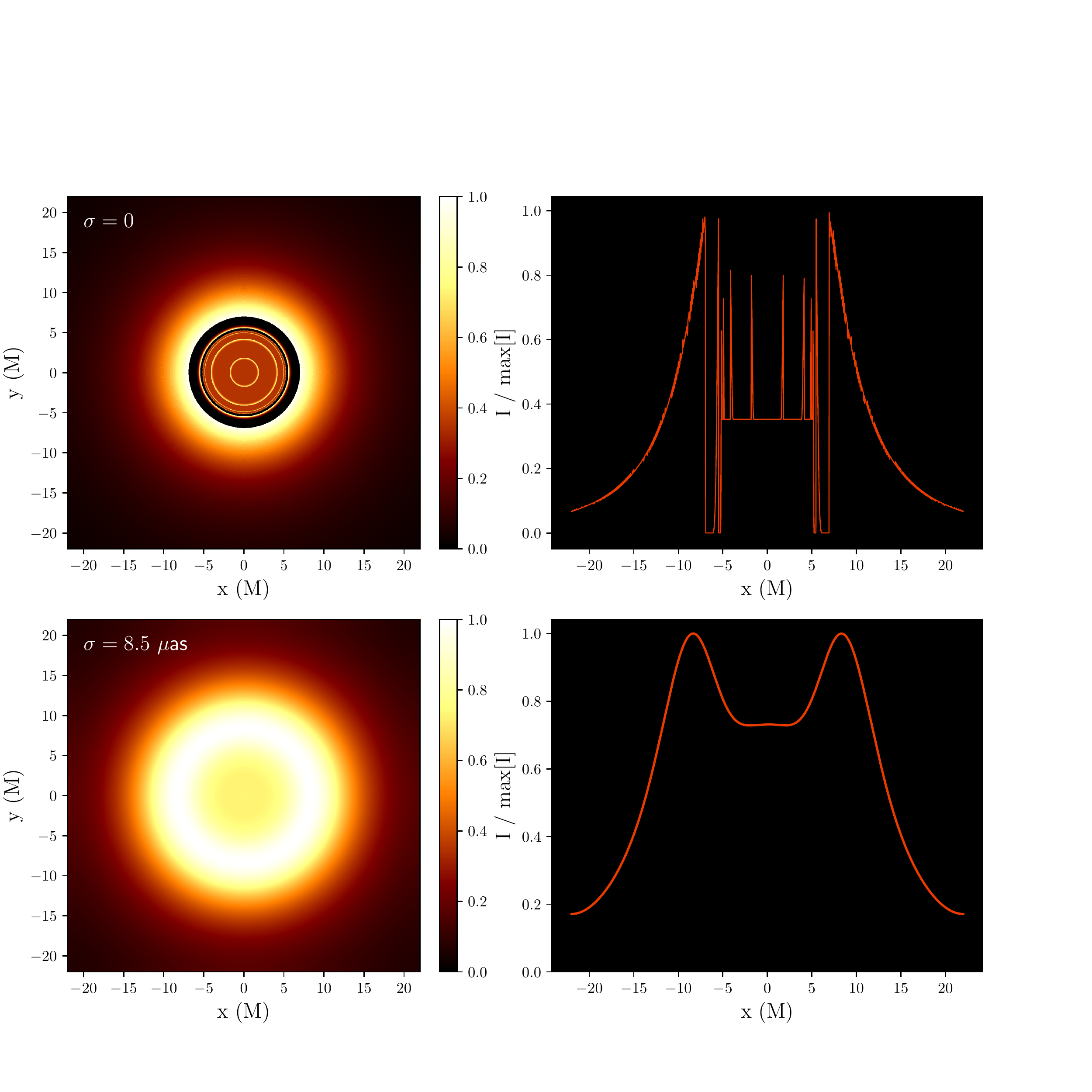}
\caption{Images of spacetimes where specular reflection takes place, but with partial absorption ($\Gamma=0.5$) and an intrinsic brightness ($\eta=10^{-2}$) included. We take $i=0^{\circ}$, $\epsilon=10^{-3}$, without filter (top row) and a Gaussian filter with the EHT angular resolution of $20\ \mu\mbox{as}$ (bottom row). We see that these values of $\eta$ change appreciably the structure of the central depression in brightness.
\label{fig:inelastic}}
\end{figure}

We treat $\eta$ as a phenomenological parameter, focusing on understanding the observability of image features for a range of its values. Specific proposals for the nature of the central object would result into specific values of $\eta$, although more complete models of the central object, as well as its interaction with matter and light, are necessary to determine these values. Also, note that $\eta$ relates the local emissivity of the surface and the total flux radiated by the accretion disc, and not the energy that the surface receives from the accretion disc. This should be taken into account when estimating the value of $\eta$ from first principles (see the discussion below). Fig.~\ref{fig:inelastic} discusses the image features associated with re-emission.

The parameter $\eta$ encodes information about the physics of the central object, which receives direct radiation from the disc but also the accreting material itself. A simple estimate for thin discs shows that the accreting material contribution, $\dot{M}$, is roughly one order of magnitude larger than the EM luminosity~\cite{1981ARA&A..19..137P}. Hence, in our simplified model in which the accretion of matter is neglected, we are underestimating $\eta$ by an order of magnitude.

If a stationary state is reached, the central object should re-radiate all the incoming energy. However, as we discussed in Section~\ref{subsec:interaction_disc_object}, a substantial fraction may not emitted as EM radiation but in other channels~\footnote{For example as positrons or high energy $\gamma$ rays, thus stimulating the calculation of accurate branching ratios, given the detection of anomalous rates from the galactic center~\cite{PAMELA:2008gwm,PhysRevLett.110.141102,DAMPE:2017fbg,HESS:2004qbs}.}. If we consider that 1\% is emitted as photons, then within our model one might expect $\eta\sim 10^{-2}$. However, the effective temperature as seen from infinity is expected to be lower than the disc's, which could lower $\eta$ by one order of magnitude. On the other hand, taking into account the accretion of matter would increase this by an order of magnitude, $\eta\gtrsim 10^{-2}$.

The above expectation holds provided there has been enough time to reach equilibrium, which depends on the value of $\epsilon$. As shown in Ref.~\cite{Cardoso:2019rvt}, this happens for $\epsilon T_{\rm salpeter}/(9.3M)\gtrsim 1$, with $T_{\rm salpeter}\sim 5 \times 10^7\,{\rm yr}$ the Salpeter time. Lowering the value of $\epsilon$ can therefore compromise reaching equilibrium in reasonable timescales. Moreover, the estimate for the timescale above assumes that no incident energy is spent in exciting the bulk degrees of freedom of the central object~\cite{Carballo-Rubio:2018jzw,Carballo-Rubio:2022imz}. Hence, our discussion of the expected values of $\eta$ above only apply to a subset of the available parameter space, but nevertheless provide a first estimation that can be used as reference in future studies, as well as combined with lower-bound constraints that discard complementary regions in parameter space~\cite{Carballo-Rubio:2018vin}.

%%%%%%%%%%%%%%%%%%%%%%%%%%%%%%%%%%%%%%%%%%%%%%%%%%%%%%%%%%%%%%%
\section{Observability of image features}
%%%%%%%%%%%%%%%%%%%%%%%%%%%%%%%%%%%%%%%%%%%%%%%%%%%%%%%%%%%%%%%

In the previous sections, we have discussed the rich features associated with specular reflection and re-emission of incident energy. These images can be understood as the idealized limit in which observational errors vanish. The different sources of errors of VLBI observations limit our ability to detect such features in practice.

\begin{figure}
\centering
\includegraphics[width=0.5\textwidth]{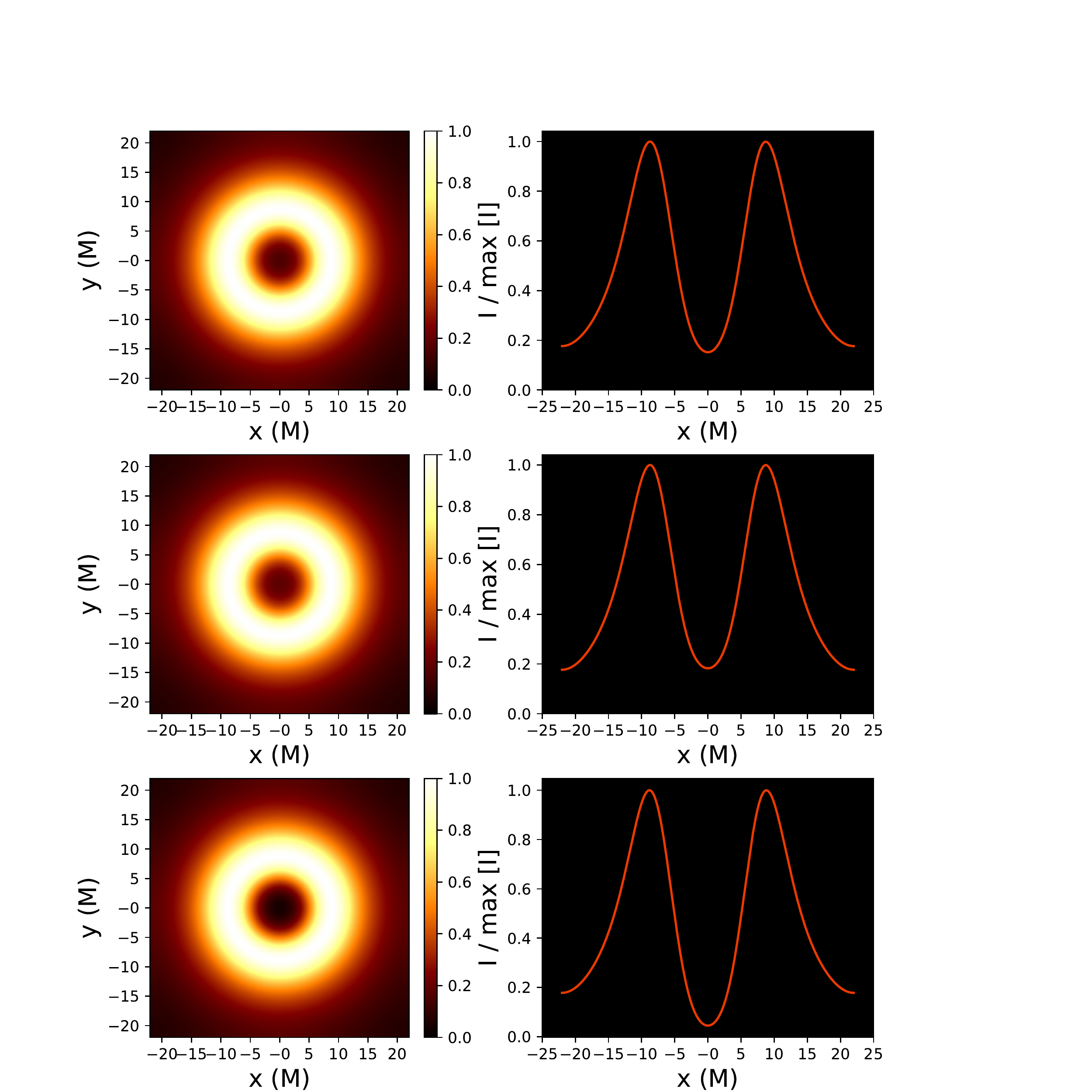}
\caption{Results of applying a Gaussian filter with the EHT angular resolution of $20\ \mu\mbox{as}$ for $i=0^{\circ}$, $\epsilon=10^{-3}$ and $\Gamma=1$, $\eta=0$ (top row), $\Gamma=0.5$, $\eta=10^{-3}$ (middle row) and $\Gamma=0$, $\eta=0$ (bottom row). We see that, even for angular resolutions that cannot resolve the internal structure, large enough dynamic ranges can allow us to distinguish between the situations in the third row and those in the first two rows. However, with this angular resolution it is generally difficult to disentangle the effects associated with the parameters $\Gamma$ and $\eta$, as the similarities between the figures in the first two rows illustrate.
\label{fig:filter_eht}}
\end{figure}

\begin{figure}
\centering
\includegraphics[width=0.5\textwidth]{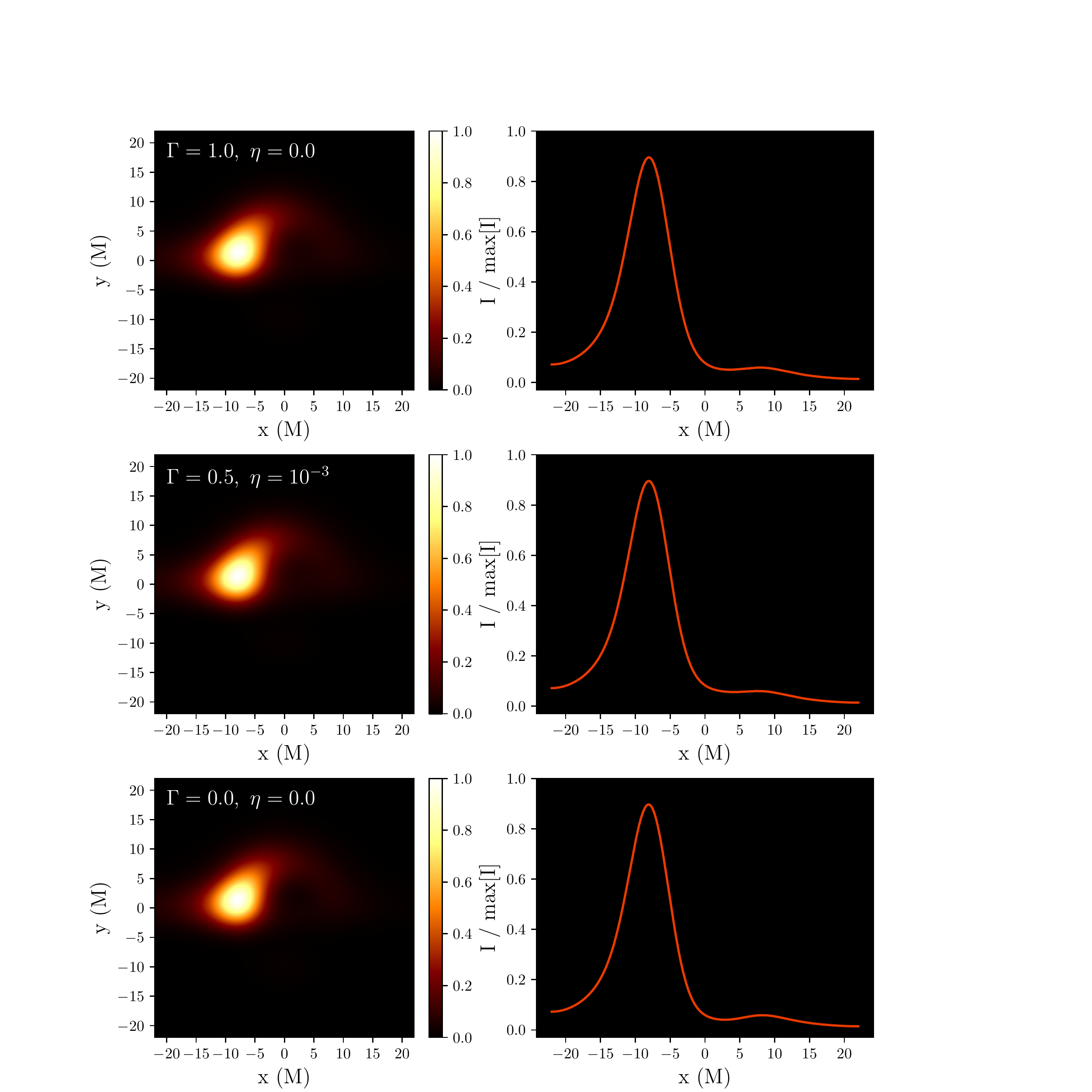}
\caption{Same as Fig.~\ref{fig:filter_eht}, but for $i=85^{\circ}$. We see that the difference between these situation shrinks, until disappearing in practice, as inclination increases.
\label{fig:filter_eht_85}}
\end{figure}

\begin{figure}
\centering
\includegraphics[width=0.5\textwidth]{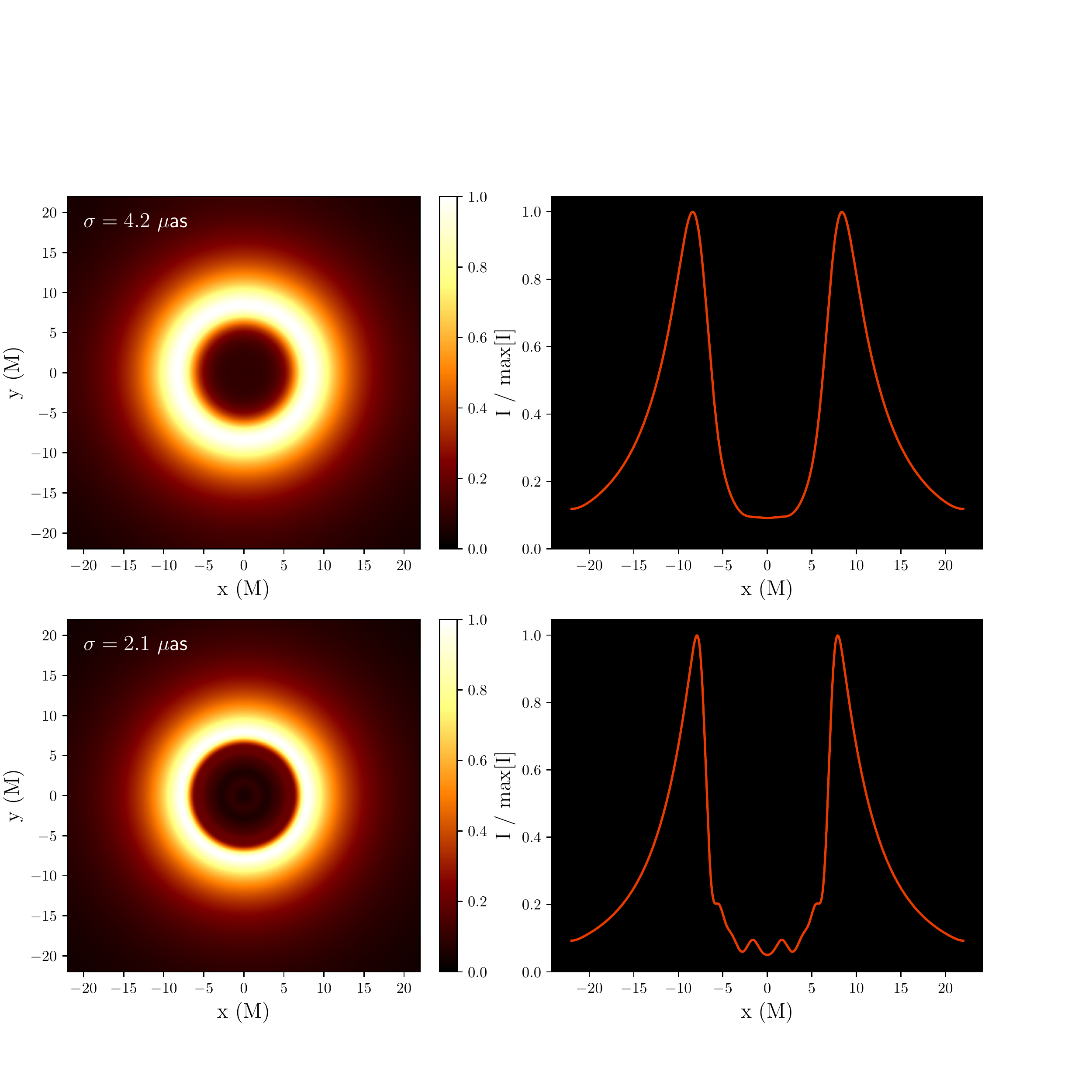}
\caption{Results of applying a Gaussian filter with an angular resolution of $10\ \mu\mbox{as}$ (top row) and $5\ \mu\mbox{as}$ (bottom row) for $i=0^{\circ}$, $\epsilon=10^{-3}$, $\Gamma=0.5$ and $\eta=10^{-3}$. For $10\ \mu\mbox{as}$, the structure of image features associated with specular reflection cannot be discerned but, if the dynamic large is large enough, it is possible to constraint the surface parameters $\Gamma$ and $\eta$. Only the most optimistic value of angular resolution ($5\ \mu\mbox{as}$) can pick up the innermost structure of the simulated image.
\label{fig:filter_ngeht}}
\end{figure}

\begin{figure}
\centering
\includegraphics[width=0.5\textwidth]{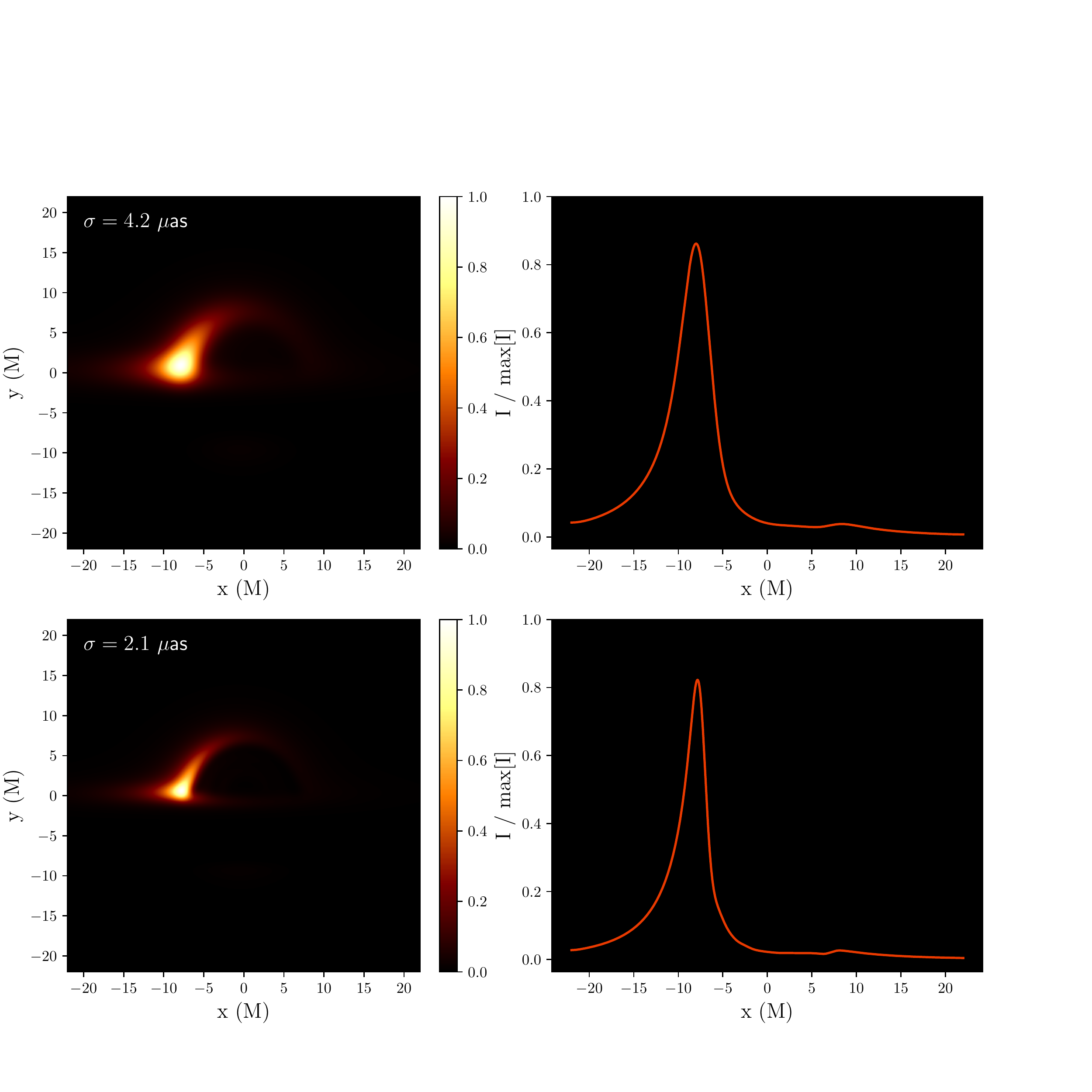}
\caption{Same as Fig.~\ref{fig:filter_ngeht}, but for $i=85^{\circ}$. The higher inclination angle makes more difficult to discern the features associated with the existence of a surface.
\label{fig:filter_ngeht_85}}
\end{figure}

The reconstruction of images according to a set of criteria, including systematic errors, but also aspects such as the array of telescopes used, is a complex problem in itself~\cite{Chael:2018oym,EventHorizonTelescope:2020eky}. However, it is standard to consider the application of a Gaussian filter as a first approximation, with the standard deviation $\sigma$ providing a measure of the angular resolution (see~\cite{Psaltis:2020cte} for a detailed discussion). We will impose that the full width at half maximum, namely $2\sqrt{2\ln2}\sigma$, equals an angular resolution of $20\ \mu\mbox{as}$ for EHT, while we consider tentative values of angular resolution of $10\ \mu\mbox{as}$ and $5\ \mu\mbox{as}$ for ngEHT. The corresponding images are shown in Figs.~\ref{fig:inelastic} -~\ref{fig:filter_ngeht_85}.

In general, we see that the image features associated both with specular reflection and re-emission are mostly filtered out for these values of angular resolution and that, generically, relatively large dynamic ranges (contrast between the brightest and dimmest points in a given image) are required to pick up these features.

For an angular resolution of $20\ \mu\mbox{as}$ and a dynamic range of $\sim10$, face-on observations would allow us to constrain $\eta\lesssim 10^{-3}$, but do not constrain in practice the specular reflection coefficient $\Gamma$ (Figs.~\ref{fig:inelastic} and~\ref{fig:filter_eht}). Higher inclination angles translate into less stringent (or virtually nonexistent) constraints (Fig.~\ref{fig:filter_eht_85}). Lowering the angular resolution to $10\ \mu\mbox{as}$ does not change much the situation (Fig.~\ref{fig:filter_ngeht}).
However, if the dynamic range increases to $\sim100$, which is within the planned capability of a future ngEHT array~\cite{Doeleman2019}, face-on observations would allow us to lower the constraint on the re-emission parameter by one order of magnitude, $\eta\lesssim 10^{-4}$, and also constrain the specular reflection coefficient $\Gamma\lesssim 10^{-1}$. Again, higher inclination angles would weaken these constraints (Fig.~\ref{fig:filter_ngeht_85}), owing to much greater image flux asymmetry and the choice of azimuthal cut made to obtain the profile cut.

%%%%%%%%%%%%%%%%%%%%%
\section{Recent complementary analyses}
%%%%%%%%%%%%%%%%%%%%%

There has recently been progress in similar, complementary directions. Naked singularities have been studied, from an imaging point of view, in Refs.~\cite{Shaikh:2018lcc} and~\cite{Guerrero:2022msp}. The latter paper focuses on the contribution of higher-order lensed images. Being naked singularities, the fate of photons that reach the core is unclear; likewise, geometries with stable light rings are expected to be unstable, thus raising questions about the viability of such models~\cite{Keir:2014oka,Cardoso:2014sna}. Similar transparent matter is assumed to compose wormhole-type geometries, in a recent study of their shadow polarization pattern~\cite{Delijski:2022jjj}. As we discussed, the interaction with the matter composing the central object is a fundamental ingredient in their appearance and imposes stringent constraints on the underlying physics.

Other situations in which the interiors of the central objects are transparent have been analyzed in~\cite{Guerrero:2022qkh} and~\cite{Eichhorn:2022bbn,Eichhorn:2022oma,Eichhorn:2022fcl}. The image features obtained in these cases are associated with photons that circle the central object but also travel across it, and are thus complementary to the ones analyzed here, as we are assuming the opposite situation in which the central object is optically thick. This consideration translates into the energy-conservation constraint $1=\kappa+\Gamma+\tilde{\Gamma}$, where $\kappa$ is the absorption coefficient, $\Gamma$ the specular reflection coefficient, and $\tilde{\Gamma}\propto\mathcal{B}$ is the dimensionless ratio between the incident and re-emitted energy. While we find the assumption of optical thickness of the central object to be physically well motivated, we understand the value of adopting an agnostic perspective and analyzing the observability of the features associated with (partially) transparent central objects. The image features analyzed in~\cite{Guerrero:2022qkh} and~\cite{Eichhorn:2022bbn,Eichhorn:2022oma,Eichhorn:2022fcl} would appear in situations in which $\kappa+\Gamma+\tilde{\Gamma}<1$, with a strength proportional to the magnitude of this deficit, and would be added to the ones analyzed here (which are present whenever $\Gamma$ or $\tilde{\Gamma}\propto\mathcal{B}$ are non-vanishing).

%%%%%%%%%%%%%%%%%%%%%
\section{Conclusions}
%%%%%%%%%%%%%%%%%%%%%

VLBI observations have the potential of unveiling the structure of BHs, and confirming whether general relativity describes it properly. However, this requires that predictions of alternative models in which this structure is substantially modified, or even replaced altogether by something else, are falsified. In this paper we have used a previously proposed general parametrization of spacetimes that, besides the perfect absorption characteristic of BHs, can describe the specular reflection and re-emission of the incident energy, which can be characteristic of exotic compact objects. We have presented a detailed analysis of the image features in the presence of a geometrically thin and optically thick accretion disc, and determined how experimental limitations impact the observability of these features.

Our main results regarding image features are the following:
\begin{itemize}
    \item Specular reflection manifests in an additional ring structure if the central object is compact enough.
    \item Re-emission manifests in a central region with uniform brightness.
\end{itemize}
These two features are nonexisting in BH spacetimes, and thus provide a concrete and novel way of testing whether a given VLBI source is a BH.

However, we have shown that, for the image dynamic range and angular resolution characteristic of EHT, and the ideal situation of $i=0^\circ$, it is possible to constraint the re-emission channel ($\eta\lesssim 10^{-3}$), but not the specular reflection channel. We have also shown how improvements in image dynamic range and angular resolution such as the ones expected to be achievable in ngEHT can noticeably change the situation, leading to more stringent constraints re-emission channel ($\eta\lesssim 10^{-4}$) and specular reflection channel ($\Gamma\lesssim 10^{-1}$), at least for $i=0^\circ$. This provides further motivation for the improvement of VLBI observations, as well as theoretical modeling aimed at extracting more precise predictions. As we argued in Section~\ref{flux_luminous}, known physics suggests that $\eta \gtrsim 10^{-2}$ once the accretion of matter is accounted for, if $\epsilon$ is greater than $10^{-10}$ (M87) or $10^{-13}$ (SgrA). Hence, ngEHT will constrain significantly the nature of the dark central object even at such small scales.

%%%%%%%%%%%%%%%%%%%%%%%%%%%%%%%
\noindent
{\bf \em Acknowledgments.} 
%%%%%%%%%%%%%%%%%%%%%%%%%%%%%%
We are indebted to Ramesh Narayan and Marek Abramowicz for valuable comments and criticisms, and to Gian Giudice for helpful discussions concerning the emission spectrum of ultracompact objects.
R.C-R. is thankful to Astrid Eichhorn, Roman Gold and Aaron Held for discussions concerning the image features in horizonless spacetimes in which photons can cross through the central object.
V.C. is a Villum Investigator and a DNRF Chair, supported by VILLUM FONDEN (grant no. 37766) and by the Danish Research Foundation. V.C. acknowledges financial support provided under the European
Union’s H2020 ERC Advanced Grant “Black holes: gravitational engines of discovery” grant agreement
no. Gravitas–101052587.
R.C-R. acknowledges financial support through a research grant (29405) from VILLUM fonden.
This project has received funding from the European Union's Horizon 2020 research and innovation programme under the Marie Sklodowska-Curie grant agreement No 101007855.
We thank FCT for financial support through Projects~No.~UIDB/00099/2020 and UIDB/04459/2020.
We acknowledge financial support provided by FCT/Portugal through grants PTDC/MAT-APL/30043/2017 and PTDC/FIS-AST/7002/2020.
Z.Y.~is supported by a United Kingdom Research and Innovation (UKRI) Stephen Hawking Fellowship, and acknowledges partial support from a Leverhulme Trust Early Career Fellowship.
This work has made use of NASA's Astrophysics Data System (ADS).

\bibliography{refs}
\end{document}